\documentclass[manuscript]{aastex}

\shorttitle{Observation of kink instability}
\shortauthors{Srivastava et al.}

\begin{document}


\title{Observation of kink instability during small B5.0 solar flare on 04 June, 2007}


\author{A.K.~Srivastava\altaffilmark{1}}
\email{aks@aries.res.in}

\author{T.V.~Zaqarashvili\altaffilmark{2,3}}
\email{teimuraz.zaqarashvili@oeaw.ac.at}

\author{Pankaj Kumar\altaffilmark{1}}
\email{pkumar@aries.res.in}

\and

\author{M.L.~Khodachenko\altaffilmark{2}}
\email{maxim.khodachenko@oeaw.ac.at}

\altaffiltext{1}{Aryabhatta Research Institute of Observational Sciences (ARIES), Nainital-263129, India}
\altaffiltext{2}{Space Research Institute, Austrian Academy of Sciences, Graz 8042, Austria}
\altaffiltext{3}{Abastumani Astrophysical Observatory at Ilia State University, Al Kazbegi ave. 2a, 0160 Tbilisi, Georgia}


\begin{abstract}
Using multi-wavelength observations of SoHO/MDI,
SOT-Hinode/blue-continuum (4504 \AA), G-band (4305 \AA), Ca II H (3968 \AA) and TRACE
171 \AA, we present the observational signature of highly
twisted
magnetic loop in AR 10960 during the period 04:43 UT-04:52 UT at 4
June, 2007. SOT-Hinode/blue-continuum (4504 \AA) observations show
that penumbral filaments of positive polarity sunspot have
counter-clock wise twist, which may be caused by the clock-wise rotation of the spot umbrae. The coronal loop, whose one footpoint is anchored in
this sunspot, shows strong right-handed twist in chromospheric SOT-Hinode/Ca II H
(3968 \AA) and coronal TRACE 171 \AA\, images. The length and the
radius of the loop are $L\sim$80 Mm and $a\sim$4.0 Mm
respectively. The distance between neighboring turns of magnetic field lines (i.e. pitch) is estimated as $\approx$ 10 Mm. The total twist angle, $\Phi\sim$12$\pi$ (estimated for the homogeneous distribution of the twist along the loop), is much larger than the Kruskal
-Shafranov instability criterion.
We detected clear double structure
of the loop top during 04:47-04:51 UT on TRACE 171 \AA \ images,
which is consistent with simulated kink instability in curved
coronal loops (T{\"o}r{\"o}k et al. 2004). We suggest, that the kink
instability of this twisted magnetic loop triggered B5.0 class solar
flare, which occurred between 04:40 UT and 04:51 UT in this active region.
\end{abstract}





\section{Introduction}
Solar coronal magnetic field has complex topology which is caused
due to photospheric motions and emergence of new magnetic flux. The
complex configurations often lead to various instability processes,
which eventually trigger solar flares and CMEs (Coronal Mass
Ejections). Kink instability is one of those processes and it is
connected to the azimuthal twist of magnetic tubes. The exact amount
of twist required to trigger the kink instability depends on various
factors including loop geometry and overlying magnetic fields (e.g.,
Hood and Priest, 1979; Lionello et al., 1998; Baty et al.,
1998; Baty, 2001; T{\"o}r{\"o}k et al., 2004; Fan and Gibson, 2003,
2004, Leka et al., 2005 and references therein). Recent observations
of kink instability accompanied with full filament eruption
(Williams et al., 2005), partial cavity eruption (Liu et al., 2007),
partial filament eruption (Liu et al. 2008), and failed filament
eruption (Alexander et al. 2006) indicate to its importance in
filament interaction with magnetic environment. Kink instability is
also found to be an efficient mechanism for solar eruptive
phenomena, e.g, triggering solar flares and CMEs ( Sakurai 1976;
Hood, 1992; T{\"o}r{\"o}k and Kliem 2005; Kliem and T{\"o}r{\"o}k,
2006 and references therein).

Although we have few observational evidences of kink instability in
various magnetic structures (e.g., coronal loops, filaments), the
theory is much more established in terms of modeling and numerical
simulations. Previous theoretical models, based on straight tube
assumption, have studied intensively various aspects of
kink instability in the solar corona including the formation of
current sheets (Baty and Heyvaerts 1996; Gerrard et al., 2001;
Gerrard and Hood, 2003; Hynes and Arber 2007) and magnetic topology
(Baty, 2000; Lionello et al., 1998). Later on, more sophisticated
models based on curved flux tube geometry, addressed the formation of current sheets (e.g. T{\"o}r{\"o}k et
al., 2004), loop response to injected twist through its footpoints
(e.g., Klimchuk, 2000; Tokman and Bellan, 2002; Aulanier et
al., 2005 and references cited there), and eruption of kink unstable
loops through overlaying arcades (e.g. T{\"o}r{\"o}k and Kliem,
2005; Fan, 2005).

In this paper, we present the observational evidence of highly
twisted coronal loop in the AR NOAA 10960 as observed on 04 June,
2007 between 04:43 UT and 04:52 UT, which probably caused B5.0 class
flare during this time.  We use observations from several different
instruments in order to cover almost whole solar atmosphere.
SOHO/MDI and SOT-Hinode/blue-continuum (4504 \AA) have been used to
observe the photospheric part of the active region. We used data
from SOT-Hinode Ca II H (3968 \AA) to study the chromospheric level.
Finally, TRACE 171 \AA \ observations are used to study the coronal
structure. In Section 2, we describe multi-wavelength observations of twist and
kink instability in AR 10960 on 04 June, 2007. In Section 3, we
present some theoretical interpretations. In the last section, we
present some discussions and conclusions.

\section{Multi-wavelength observations of kink instability in AR 10960\label{sec:obs}}
The active region NOAA 10960 has been observed by various
spacecrafts on 04 June, 2007. The active region was located nearby
the eastern limb of the southern hemisphere near the equatorial
plane at S09E50 with $\beta\gamma\delta$ field configuration of the sunspot group. This
active region produced ten M-class flares during its journey over
the whole solar disk. However, this active region was poor CME
originator, and only two from all M-class flares were associated
with CMEs. The active region has produced a
confined M8.9/3B flare on 04 June, 2007 without triggering CME.
The flare has an impulsive rise phase for a short duration between
05:06 UT and 05:13 UT and a gradual decay for a long time nearly up
to 06:45 UT.
Kumar et al. (2009) have recently described a detailed multiwavelength
view of this M8.9/3B class solar flare, and concluded that the positive flux emergence and the penumbral filaments
loss of the associated sunspot and the activation of the several twisted flux ropes
in and around the flare site, may be the key candidates for the occurance of this flare.
However, a small B5.0 class flare has also been observed during
04:40-04:51 UT, which seems to be a precursor for the M8.9 flare.
This B-class, small and confined flare was also evident in the GOES
soft X-ray profile during 04:40 UT--04:51 UT of 04 June, 2007.

Below we explore the relation between this B-class small
flare, which may have triggered the M-class flare later,
and the magnetic structure of the active region. During the
flaring time, we detected the strong helical twist in the coronal
loop, which was associated to the flare. Figure 1 shows the coronal
image of AR 10960 as observed from Transition Region and Coronal
Explorer (TRACE) in 171 \AA\ line on 04:48:55 UT at 04 June, 2007.
The TRACE field of view is overlaid by co-aligned MDI (Michelson
Doppler Imager) contours showing positive (white) and negative
(black) magnetic polarities. One footpoint of the coronal loop is
associated with a small positive magnetic polarity spot (above which
the helical twist and simultaneous brightening has been observed
during 04:43 UT and 04:52 UT). Another footpoint is probably
anchored in a small leading spot with negative polarity in the same
active region.

High-resolution filtergrams of NOAA 10960 at photospheric and
chromospheric levels were obtained by 50 cm Solar Optical Telescope
(SOT) onboard Hinode spacecraft at 04 June, 2007. We use the
SOT/blue-continuum (4504 \AA) , G-band (4305 \AA) photospheric and SOT/Ca II H (3968
\AA) chromospheric image data at a cadence of $\sim$1 minute, with a
spatial resolution of 0.1$''$ per pixel (Tsuneta et al., 2008). The
data are calibrated and analysed using standard routines in the
SolarSoft (SSW) package.

Figure 2 displays the partial field of view of TRACE 171 \AA\ image
on 04:48:15 UT at 04 June 2007, which shows the coronal loop segment
with activated helical twist. The TRACE temporal image data of 1 minute cadence is
calibrated and also co-aligned using Solar-Soft routines. The co-aligned SOT G-band (left panel)
and Ca II (right panel) contours are overlayed on TRACE 171 \AA \
image. The G-band contour shows the position of sunspot which
probably is the source of helical twist, while the Ca II contour
shows the chromospheric part of the loop segment.The solar loops associated with active regions of the southern hemisphere,
usually show the predominant right-handed twist according to the
Hemisphere Helicity Rule (HHR) (Tian et al., 2002, 2005 and references cited there).
The coronal loop shown in the Figure 2 also exhibits a right-handed twist and thereby follows the HHR.

Figure 3 shows the time series of SOT/blue-continuum (4504 \AA)
during 04:36 UT and 04:52 UT at an intervals of 4-5 min on 04 June, 2007 (lower panel).
This small positive polarity spot, where the left footpoint of twisted coronal loop is anchored, is a small part of a big active region AR10960 (upper panel). The active region was associated with the descending phase of the solar cycle 23. It obeys both, the Hale-Nicholson Law (Hale and Nicholson, 1925) and the Joy's Law (Hale et al., 1919).
Bipolar active regions are believed to be formed due to
rising of $\Omega$-shaped magnetic flux tubes through the convection
zone (Babcock, 1961, Zwaan, 1985). The axes of bipolar active
regions are then tilted towards the equator due to the Coriolis
force resulting in a counter-clockwise rotation in the southern
hemisphere and vice-versa (e.g., Wang \& Sheeley, 1989;
Howard, 1991; Tian et al., 2001; 2002; 2003, 2005, Zhang
\& Tian 2005 and
references therein). Numerical simulations also show that the newly emerged magnetic
tubes are supposed to be twisted during the rising phase through the
convection zone (Moreno-Insertis and Emonet 1996; Archontis et al.
2004). On the other hand, Sturrock \& Woodbury (1967) proposed that the photospheric latitudinal differential rotation may lead to a clockwise (counterclockwise) rotation of sunspots about their axes in the southern (northern) hemisphere. Observed rotation of sunspots (Brown et
al. 2003, Yan and Qu 2007) may lead to twist in and above active regions.
The high-resolution morphology on Figure 3 shows that the penumbral filaments of the
spot are highly twisted and sheared in the counter-clockwise direction. In order to be sure about the emergence of new magnetic flux during B5.0 flare in AR 10960 (04:40 UT--04:51 UT), the high cadence magnetograms (e.g.,SOT/Hinode magnetograms) are needed. Unfortunately,
the SOT magnetograms are unavailable for this period. We examined 96-min cadence
SoHO/MDI magnetograms, available at 03:15 UT and 04:51 UT, and found no new flux emergence in and around
the positive polarity sunspot. Therefore, under the observational
limitations, it is most likely that the clock-wise rotation of the spot umbrae is responsible for the counter clock-wise twist of penumbral filaments. Indeed, some evidence of insignificant clock-wise rotation of the umbral part can be seen from 04:36 UT and later (see lower panel, Figure 3). The rotation is more pronounced in SOT/blue-continuum movie. Unfortunately, there are no observational data before the time, therefore we are not completely sure about the rotation and its rate.

Rotation and twisting of sunspots at the photospheric level
eventually lead to the twisting of magnetic field lines in higher
regions. Therefore, the rotation of the spot probably play a role
in the generation of right-handed twist in associated coronal loop which is consistent with many previous findings
(e.g. Bao et al. 2002; Brown et al. 2003, Tian et al. 2008 and references therein). Tian et al. (2008) have found the
significant counter-clockwise rotation in the AR sunspots of northern hemisphere during the same solar cycle which generates
the left-handed twist in the associated loop system.
Figure 4 displays the time series of SOT/Ca II H (3968 \AA)
during 04:43 UT and 04:52 UT at 04 June 2007.
The time series shows the clear twist just above the spot and
consequent activation of the helical twist in the chromospheric part
of the coronal loop during B5.0 class flare (i.e. between 04:40
UT--04:51 UT). The brightening of chromospheric plasma is also seen
during the time interval, which may show the evidence of heating.

Helical twist is more clearly seen in the coronal part of AR 10960.
Figure 5 shows the time series of TRACE 171 \AA \ in the same active
region and during the same time interval 04:43-04:52 UT.
The plasma brightening during the B5 flare is clearly seen.
The brightening reveals clear helical twist of the coronal loop
which connects two different polarities in the active region (see
Figure 1).
The writhe of the coronal loop is not clearly identified,
therefore we can not say whether the loop has S- or reverse
S-structure. However, the right-handed twist can be easily identified between
04:45 and 04:52 UT, especially in its left footpoint. Thus, the loop
was strongly twisted during the B5.0 flare (04:40-04:51 UT). The loop footpoint is only seen during the brightening
phase i.e. during 04:40 UT--04:51 UT, therefore we could not trace how the twist was formed. It seems that the loop was already twisted, but the brightening just made it observable.

Especially interesting event is observed during 04:47-04:51 UT on
TRACE 171 \AA \ images (Figure 5). It seems that the loop top is
split into two parts and shows double structure during this time
(the double structure is clearly seen on Figure 1 as well). We have also
analyzed XRT/Hinode Ti-Poly temporal image data of the same active region.
We use the standard methods to calibrate the data and correct the orbital
variations to see the co-spatial X-ray brightening of the flare in this loop (Shimzu et al., 2007).
Sudden
brightening and enhancement of soft X-ray flux in the loop has also
been observed around 04:49 UT by XRT/Hinode.
Thus, the double structure of loop top in 171 \AA \ coincides
to the enhancement of soft X-ray flux.

Recently T{\"o}r{\"o}k et al. (2004) performed numerical simulations
of kink instability of twisted coronal loops and show that the kink
perturbations with a rising loop apex lead to the formation of a
current sheet below the apex, which does not occur in the
cylindrical approximation. The double structure of coronal loop at
its top is seen in the numerical simulations (see Figure 3 in
T{\"o}r{\"o}k et al. 2004), which is quiet similar to that we
observe during 04:47-04:51 UT.
Therefore, we find the most likely evidence of the kink instability in the twisted coronal
loop, which triggers the small B5.0 class flare during 04:40 UT--04:51 UT.

In the next section we estimate the loop parameters, the value of
twist and make some theoretical suggestions.

\section{Theoretical estimations \label{sec:theory}}
%

Most clear image of twisted loop in TRACE 171 \AA \ series is seen
at 04:49:36 UT, therefore we use this frame for estimation of loop
parameters.

The projected distance between loop footpoints is $\sim$ 40 Mm.
However, the real distance can be increased up to 50 Mm taking into
account the projection effects. Using the ideal semi-torus form we
may estimate the loop real length as
\begin{equation}
L=\pi R \approx 80 \,\,Mm,
\end{equation}
where $R$ is the big radius of torus. The loop small radius at the
middle of left side can be estimated as $a \approx 4\,\, Mm$.

An important parameter of straight twisted tubes is
\begin{equation}
p ={{aB_z}\over B_{\phi}},
\end{equation}
where $B_z$ and $B_{\phi}$ are the longitudinal and azimuthal
components of magnetic field, and $2\pi p$ is called as {\it pitch}.
Then the pitch can be expressed by loop length and the number of
turns over the tube length ($N_{twist}$) as
\begin{equation}
2\pi p= {L\over N_{twist}}.
\end{equation}

The total twist angle is then

\begin{equation}
\Phi= 2\pi N_{twist}= {{LB_{\phi}}\over {a B_z}}.
\end{equation}

At least, 3 different turns are seen along the left half length of
the loop at 04:49:36 UT (see Fig. 5). The mean pitch, i.e. the distance between neighboring turns of magnetic field lines, is estimated as $2\pi p \approx$ 10 Mm, while the total twist angle for the left half of the loop is roughly
\begin{equation}
\Phi \approx 6.0 \pi.
\end{equation}
The right footpoint is not
seen clearly in the TRACE images, therefore we can not estimate the number of turns
there. Some helicity imbalance may exist between the two ends of the loop (Tian and Alexander, 2009; Fan et al., 2009).
This is also evident from the Hinode/SOT WB full FOV image (Figure 3), where the opposite polarity sunspots seem to have
imbalanced morphology. However, this asymmetric helicity (if any) does not
mean that the twist will be inhomogeneous along the loop. The initial asymmetric distribution of twist along the loop can be smoothed out over Alfv\'en time (Fan et al., 2009). Our observed loop with the length of 80 Mm and average coronal Alfv\'en speed of 1000
km s$^{-1}$ implies the Alfv\'en time as $t=L/V_A \sim$ 80 s. Thus, any asymmetry in the twist distribution will
be smoothed out within $\sim$ 80 s. In other words, the rotation of one footpoint may cause the twisting of whole loop, even if the spot, where the second footpoint is anchored, does not rotate. Therefore, we may consider the quasi-symmetric twist along the whole
loop during 04:40 UT--04:51 UT when B5.0 flare was triggered. Then 3 more turns along the right half length of the loop are expected, which give the total twist angle for the whole loop as
\begin{equation}
\Phi \approx 12 \pi.
\end{equation}

This is a rather strong twist. However, even if we consider the lower bound of total twist angle, 6$\pi$ (estimated from the left half part of the loop), the loop could be unstable to kink instability, which eventually may lead to observed
B-class flare.

Kink instability is caused due to the growth of $m=1$ mode, which
displaces the tube axis. General properties of classical kink
instability are well studied. Kruskal -Shafranov instability
criterion yields $\Phi > 2\pi$. Line tying of loop footpoints at the
photospheric level generally increases the critical twist angle and
the instability threshold becomes $\Phi > 2.5 \pi$ (Hood and Priest
1979). However, for the large aspect ratio (i.e. the ratio between
loop length and radius), the critical twist angle increases further
(Baty 2001). The approximate aspect ratio of our loop is $L/a
\approx$ 20, which is quiet large. Normal mode analysis in the thin
tube approximation (i.e. for large aspect ratio) gives the
instability criterion as $a>2p$ (Dungey and Loughhead, 1956, Bennett
et al. 1999), which yields also large twist angle. In our loop
parameters, this gives the critical twist angle of $\Phi \approx
12\pi$, which is very close to the angle estimated from the homogeneous twist. Longitudinal flow along twisted magnetic tube may reduce the critical twist angle, thereby enhances the probability of the kink instability (Zaqarashvili et al. 2010). However, no clear evidence of flow is detected in our observations.

Numerical simulations of T{\"o}r{\"o}k et al. (2004) have been
performed for a curved loop taking into account the line-tying
conditions. They used the twisted loop model of Titov and Demouline
(1999) and show that the loop is subject to ideal kink instability.
They estimated the critical twist angle for different parameters of
coronal loops. T{\"o}r{\"o}k et al. (2004) solved the case of
$R/a=5$ and showed that the critical twist angle is $\Phi_c \approx
3.5$ for $R$=110,000 Mm, although these parameters are much larger
comparing to our loop. On the other hand, the critical twist angle
probably increases for smaller $a$ (Baty 2001).
Unfortunately, T{\"o}r{\"o}k et al. (2004) did not estimate a
critical twist angle for loop parameters, which are close to our
loop. However, general behaviour of the twisted loop could be
similar to what they have shown in their paper. Namely, the double
structure of loop top (Figure 3 of their paper) seems quiet similar
to what we have observed here (Figure 5).

\section{Discussion and conclusions}
Using multiwavelength observations of SoHO/MDI,
SOT-Hinode/blue-continuum (4504 \AA), G-band (4305 \AA), Ca II H (3968 \AA) and TRACE
171 \AA , we find the observational signature of highly twisted
coronal loop in AR 10960 during the period of 04:43 UT-04:52 UT at
04 June, 2007. This twisted loop was probably unstable to the kink
instability, which triggered small B-class flare during the same
time. SOT-Hinode/blue-continuum (4504 \AA) images show that the
penumbral filaments of small positive polarity sunspot have
counter-clock wise twist around its center (Figure 3). The twist is probably caused by the clockwise
rotation of the sunspot umbrae as expected in the southern hemisphere.  The observed coronal loop whose
one footpoint is anchored in this spot, shows strong right-handed
twist in chromospheric (Ca II H 3968 \AA \ , Figure 4) and
coronal (TRACE 171 \AA \ , Figure 5) images. We estimated the loop
length and radius as $\sim$ 80 and $\sim$ 4 Mm respectively. From
TRACE 171 \AA\ time sequence we estimate the pitch of twisted loop as $\sim$ 10 Mm. The twist is assumed to be homogeneous along the whole loop as any asymmetry can be quickly
smoothed out over the Alfv\'enic time, which is estimated as $\sim$80 s for our loop. The total twist angle for the whole loop is $\Phi$$\sim$12$\pi$ in the case of homogeneous distribution.
This is much larger than the Kruskal-Shafranov
kink instability criterion, and approximately equal to the more
conservative thin-tube estimate for the kink instability.
The loop top
shows clear double structure during 04:47-04:51 UT, which was
accompanied by sudden enhancement of soft X-ray flux around 04:49 UT
observed by XRT/Hinode.

The strong twist of the coronal loop probably leads to the kink
instability, although no clear displacement of the loop axis or
sigmoid structure have been detected during the flaring time.
However, the observed double structure of the loop top probably
indicates to the current sheet induced by the kink instability as it
was suggested by numerical simulations (T{\"o}r{\"o}k et al. 2004).
This current sheet probably triggered the B-class flare via
reconnection.

Instability of internal kink mode (Arber et al. 1999; Haynes and
Arber, 2007), where the kink structure is not apparent from the
global field shape of the active region, may be considered as a
triggering mechanism for the B-class flare. Haynes and Arber (2007)
considered the example of short coronal loop ($L$= 10 Mm) with zero
net current. This means that the magnetic twist changes the sign
along the loop radius, which causes the confinement of unstable kink
mode in space. This process has also been shown by Arber et al.
(1999) to release sufficient energy to cause a transient brightening
of confined loops. However, there are some questions related to the simulation of Haynes and Arber (2007). First, our observed loop is
longer compared with the simulated one by a factor of 8. Therefore,
it is unclear if the same mechanism may work for longer loops.
Second, it is unclear why the twist should have opposite signs along
the tube radius. Haynes and Arber (2007) considered photospheric
motion as a generator of this configuration, however no clear
process has been suggested.

Twisted coronal loops can be also subject to the sausage
instability. Linear sausage pinch instability occurs when
$B^2_{\phi}> 2 B^2_z$ (Aschwanden 2004). It gives the critical twist
angle of $\sim 9 \pi$ in our loop parameters. Nonlinear analysis of
Zaqarashvili et al. (1997) shows even smaller critical twist angle.
The observed twist in our loop is much larger. However, the sausage
pinch does not lead to observed double structure at the loop top,
which rules out the occurrence of sausage instability in the loop.

It is quiet possible that the observed twist of coronal loop corresponds to the
helical structure formed after the kink instability. Then, the radius of the loop will be
bit smaller than $\sim 4$ Mm. However, this means that the wave length of unstable mode is $\sim 15$ Mm (i.e. twice the distance between neighboring bright areas along the loop, see Figure 5), which is much less comparing to the loop length. It means that the higher order harmonic rather than fundamental one was unstable to the kink instability. This unexpected fact needs special justification, which may require detailed numerical simulations.

It should be mentioned, that the small B5.0 flare was precursor for
stronger M-class flare, which occurred just 15 min later at 05:06
UT. It is possible that the magnetic field reconfiguration due to the
small B5.0 flare caused global kink instability with consequent
reconnection and global energy release. It is interesting to check
whether the big flares are usually preceded with smaller energy
release. Liu et al. (2003) and Gary and Moore (2004) have observed recurrent flare activity of the active region AR 10030
in the northern hemisphere on 15 July 2002, which was accompanied by CME. The large flare/CME was preceded by a small flare and the authors supposed the rise of helical flux rope as the source for the activity. Their observations show strongly twisted loop in TRACE images (see Fig. 1 in Gary and Moore, 2004), which arose upwards and disappeared during short time interval ($\sim$ 50 s). One may suggest that the strong twist of our loop may be also formed due to the similar rise of magnetic flux rope. However, there are significant differences between the observations. First, no new magnetic flux emergence is seen at the photospheric level in our case. And second, our twisted loop stays unchanged during longer time interval (at least 5 min) and no sign of upward motion is detected. Therefore, we think that the rise of helical flux rope is not relevant to our observations,
which probably suggests at least two different triggering mechanisms for solar flares. The similarity between the two events is that the small flares seem to be the pre-cursors of large flares. On the other hand, the flare in the active region AR 10030 on 15 July 2002 was accompanied by CME (Liu et al. 2003, Gary and Moore 2004), while the flare in AR 10960 on 04 June 2007 was not (Kumar et al., 2010).
Therefore, we suggest that the large flares accompanied with energetic CMEs can be triggered by flux-rope eruption with significant
changes in the photospheric fields, while the moderate flares without CME are triggered by some instabilities (e.g., kink instability). More statistical study is required to make a firm conclusion.

In conclusion, we observe the strong twist of coronal loop in AR
10960 during small B5.0 flare between 04:43 UT--04:52 UT at 04 June,
2007. The loop top shows clear double structure, which is consistent
with simulated kink instability of curved coronal loop
(T{\"o}r{\"o}k et al. 2004). We suggest that the current sheet
formed at the loop top due to the kink instability was the reason
for the B5.0 flare.

\acknowledgments

This work is supported by the grant of a joint Indo-Russian (INT/RFBR/P-38)
DST-RFBR project, the Austrian Fond
zur F\"orderung der Wissenschaftlichen Forschung (project
P21197-N16) and the Georgian National Science
Foundation grant GNSF/ST09/4-310. Hinode is a Japanese
mission developed and launched by ISAS/JAXA, with NAOJ as domestic
partner and NASA and STFC (UK) as international partners. It is
operated by these agencies in co-operation with ESA and NSC
(Norway). We also acknowledge MDI/SoHO and TRACE observations used
in this study. We thank the anonymous referee for constructive
suggestions which improved the manuscript considerably.

\clearpage


\begin{figure}
\centering
\includegraphics[width=15.0cm]{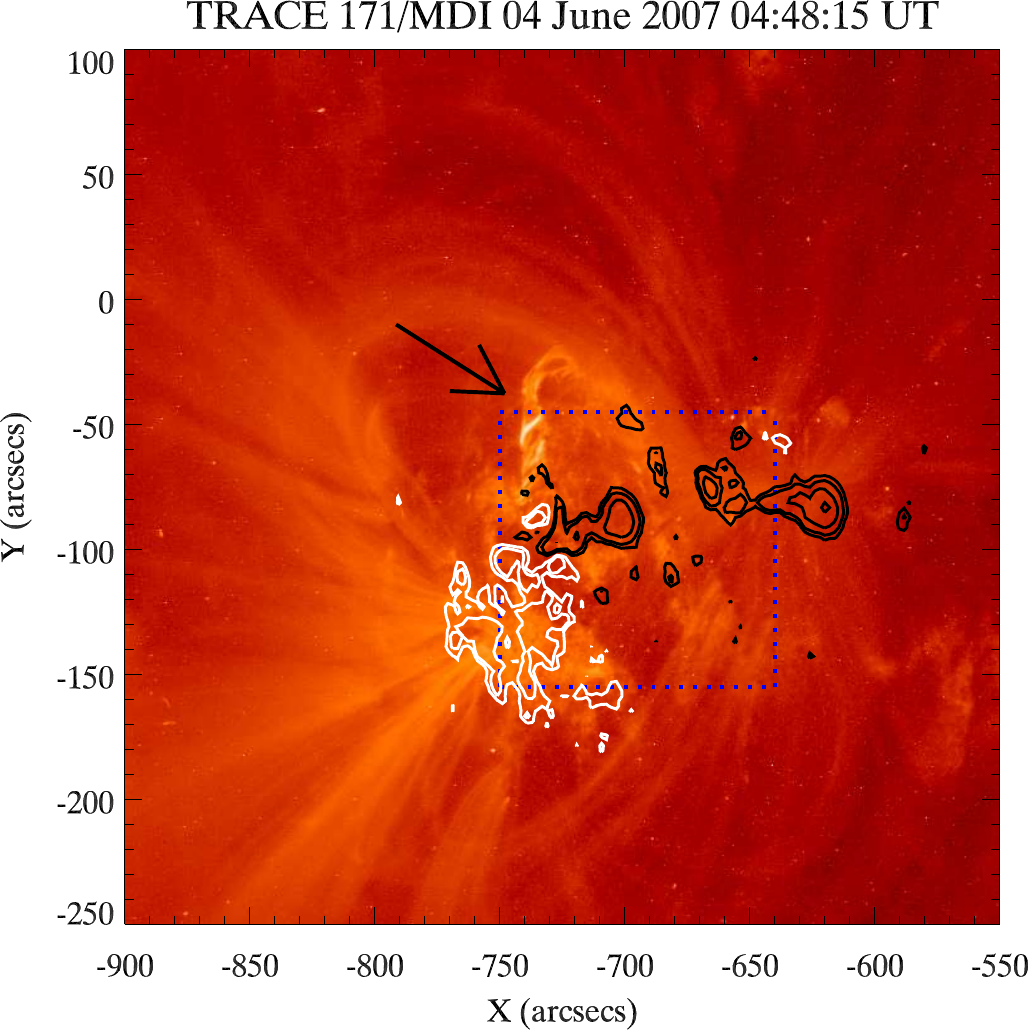}
\caption{Co-aligned MDI contours overlaid on TRACE 171 \AA\
image of the active region AR 10960 on 04:48:15 UT at 4 June, 2007.
White (black) contours show the positive (negative) polarity of photospheric magnetic field.
The loop with helical twist is shown by the black arrow, while SOT-FOV by blue-dotted square. The left footpoint of the loop is anchored in the small positive polarity spot at (X,Y)=(-735,-90), while the right footpoint is probably anchored in the negative polarity spot at (X,Y)=(-700,-50).
}
\label{fig1}
\end{figure}
\clearpage
\begin{figure}
\centering
\includegraphics[width=6.2cm]{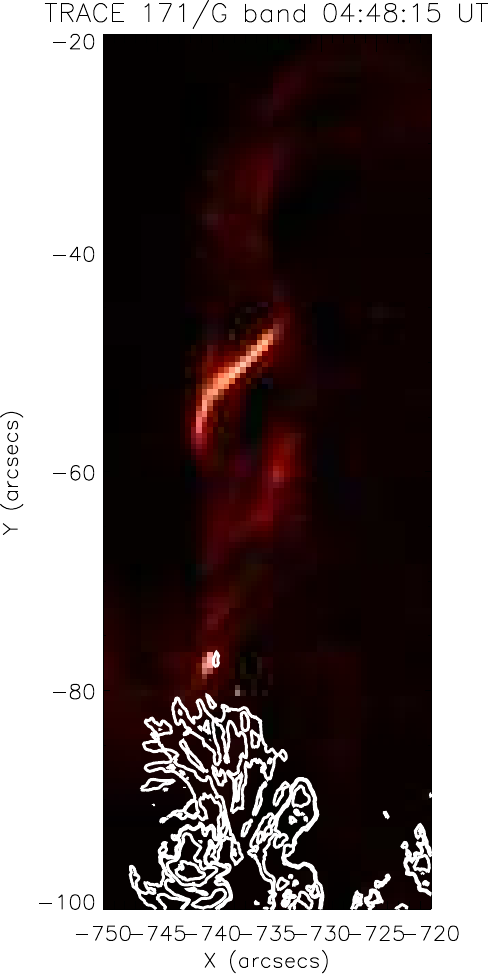}
\includegraphics[width=6.0cm]{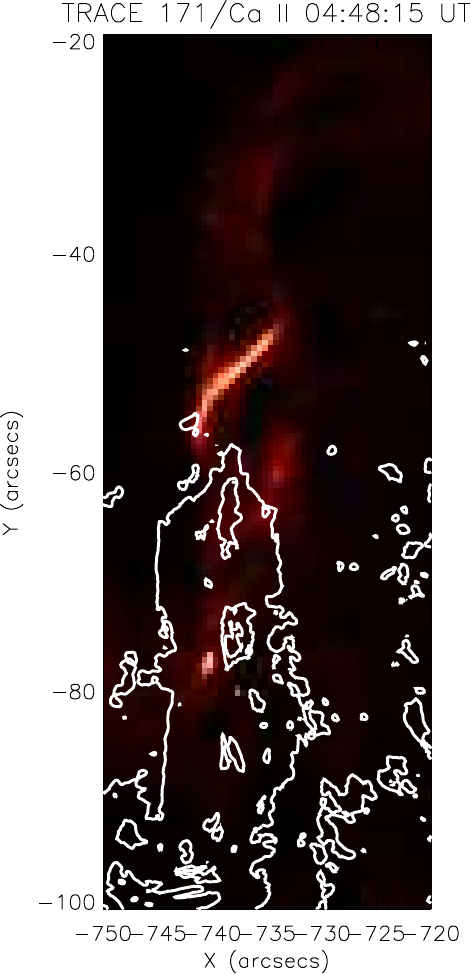}
\caption{Partial FOV of TRACE 171 \AA\ image on 04:48:15 UT at 04 June 2007, which shows
the coronal loop segment with strong helical twist. The co-aligned SOT G-band (left panel)
and Ca II (right panel) contours are overlayed on TRACE 171 \AA \ image, which show
the sunspot position and the chromospheric part of the loop respectively.}
\label{fig2}
\end{figure}

\clearpage
\begin{figure*}
\centering
{\includegraphics[width=12cm]{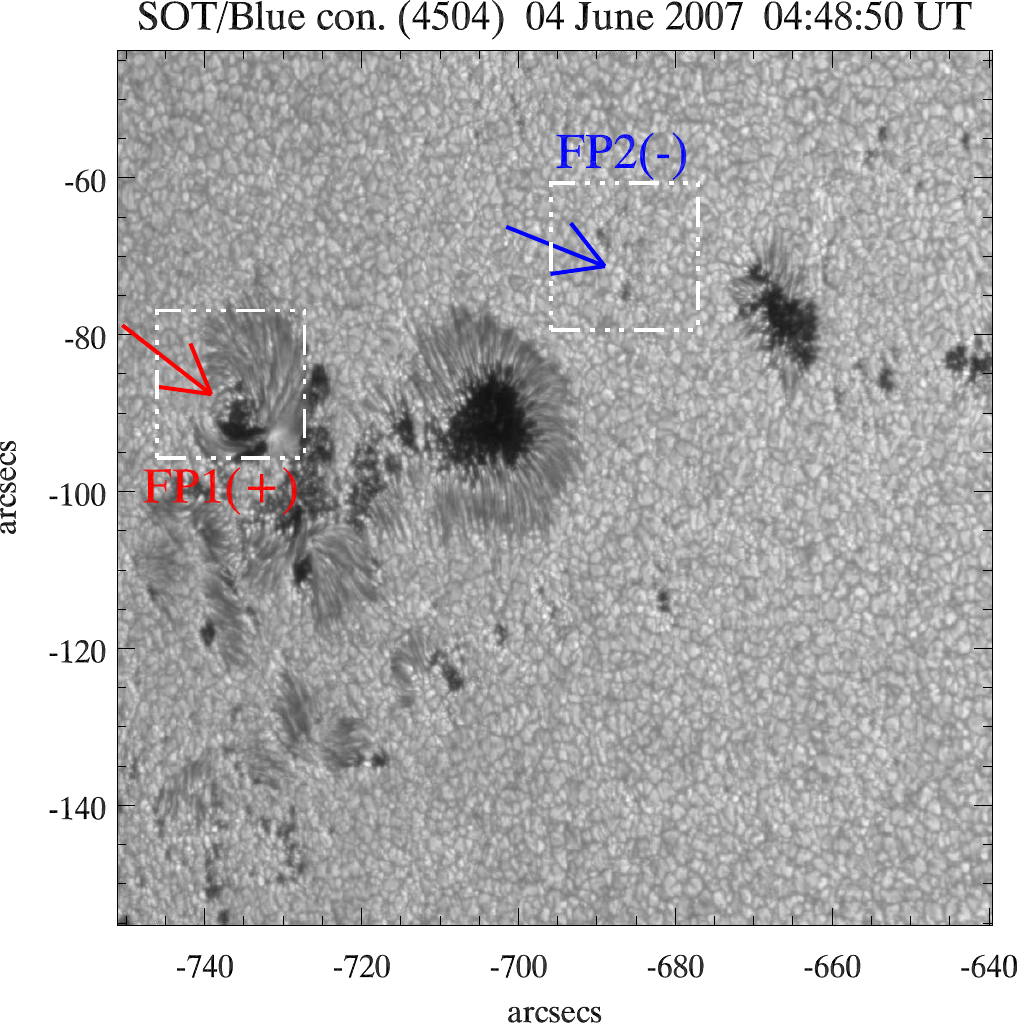}
}

{
\includegraphics[width=4cm]{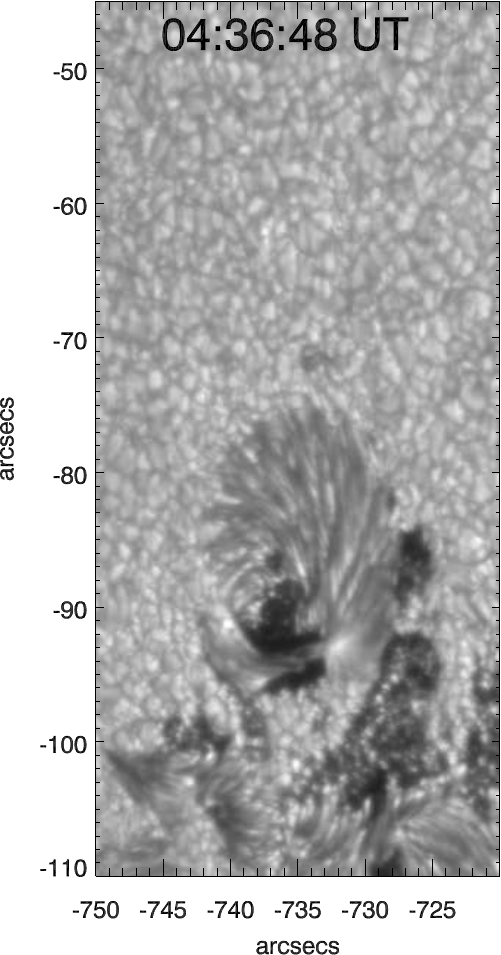}
\includegraphics[width=3.25cm]{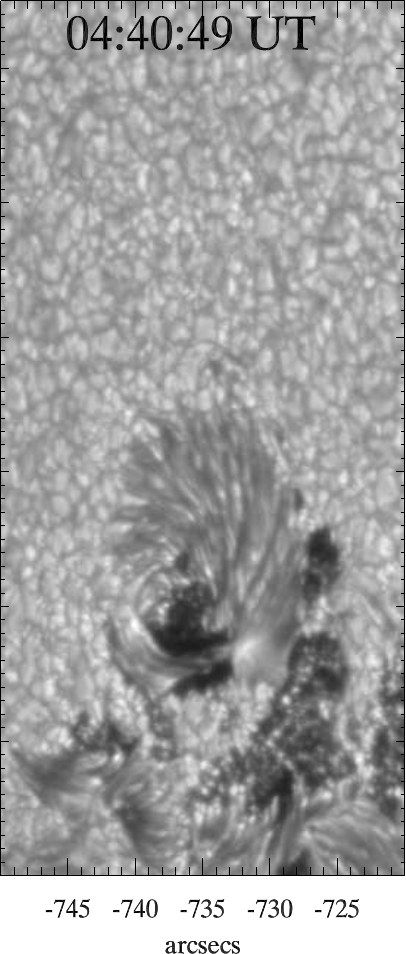}
\includegraphics[width=3.25cm]{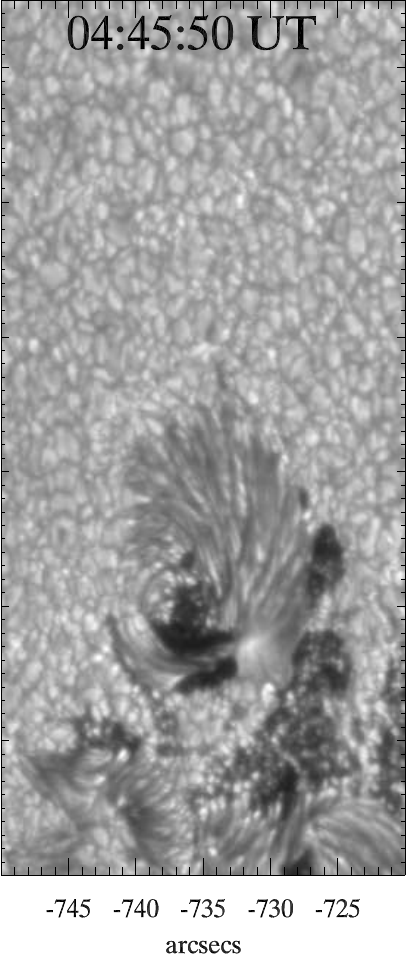}
\includegraphics[width=3.415cm]{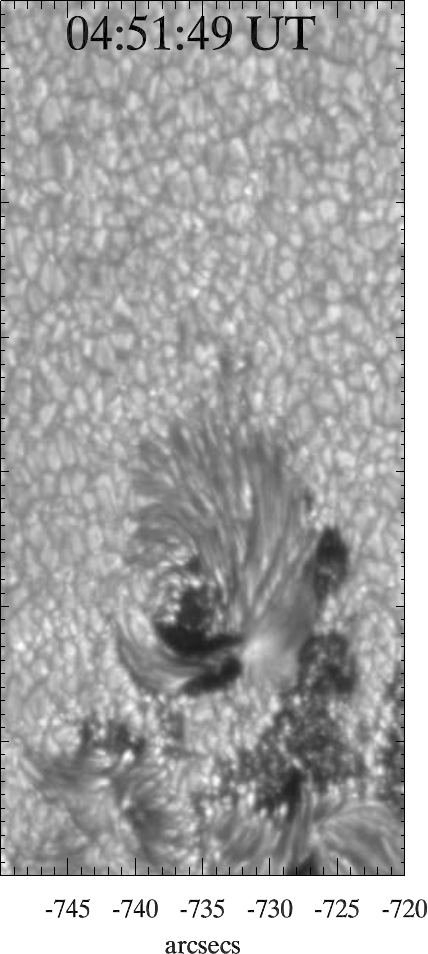}
}
\caption{The full FOV of AR 10960 in SOT/blue-continuum (4504 \AA) on 04:48:50 UT, which shows the two opposite polarity spots [FP1(+) and FP2(-)]in which respectively the left and right footpoints are anchored (top panel).
Partial FOV of AR 10960 in SOT/blue-continuum (4504 \AA ) during 04:36 UT and 04:52 UT at 04 June, 2007 at each $\sim$
3-4 min interval. The time sequence shows the high-resolution morphology of the spot, where the left footpoint of helically twisted coronal loop is anchored. The spot is only small part of big active region - AR 10960 (bottom panels). }
\label{fig3}
\end{figure*}
\clearpage
\begin{figure*}
\centering
{
\hspace{0.01cm}
\includegraphics[width=3.1cm]{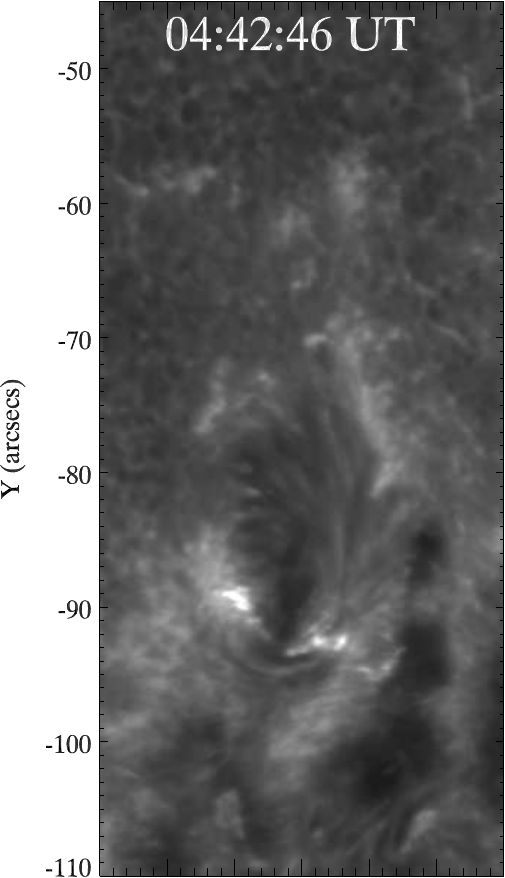}
\hspace{0.12cm}
\includegraphics[width=2.5cm]{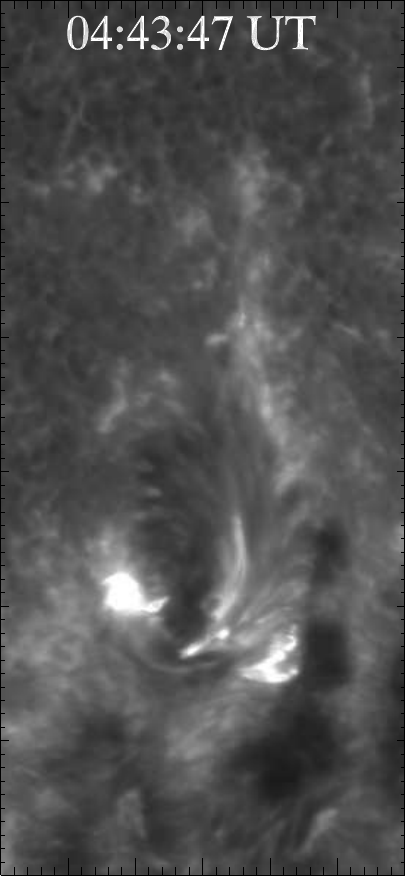}
\hspace{0.12cm}
\includegraphics[width=2.5cm]{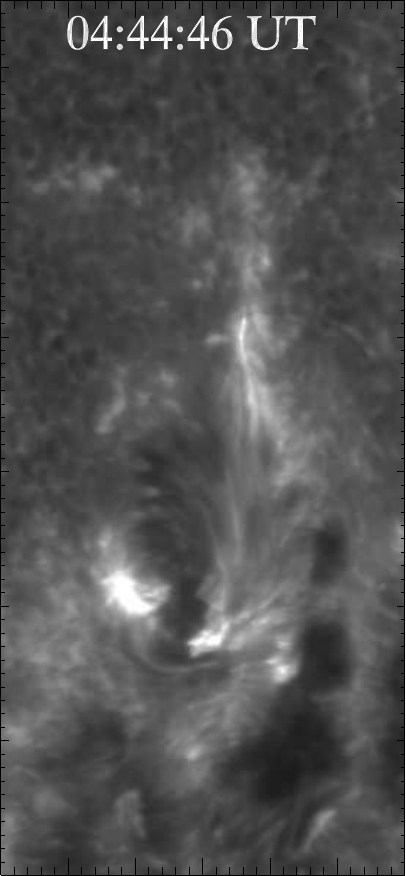}
\hspace{0.12cm}
\includegraphics[width=2.5cm]{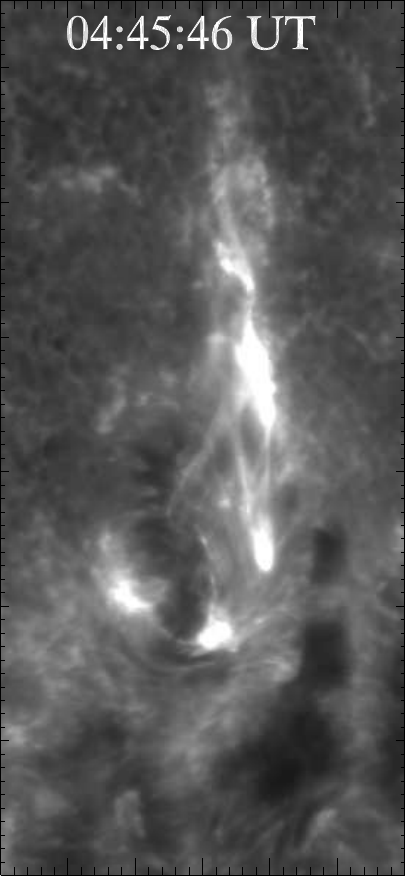}
\hspace{0.12cm}
\includegraphics[width=2.5cm]{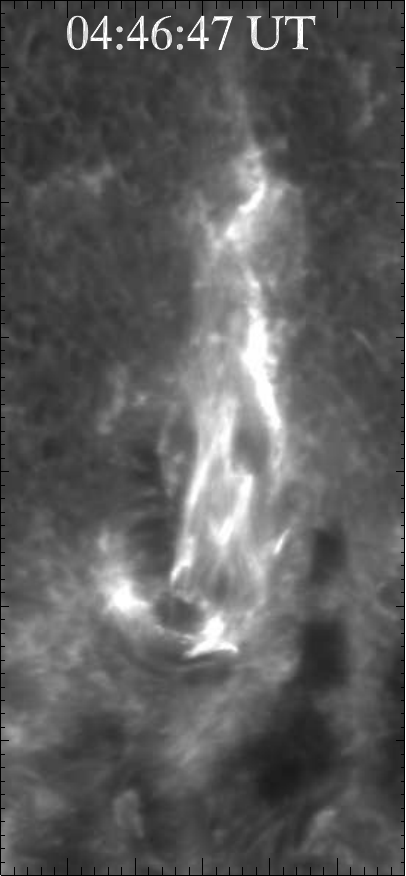}

\hspace{0.2cm}
\includegraphics[width=3.3cm]{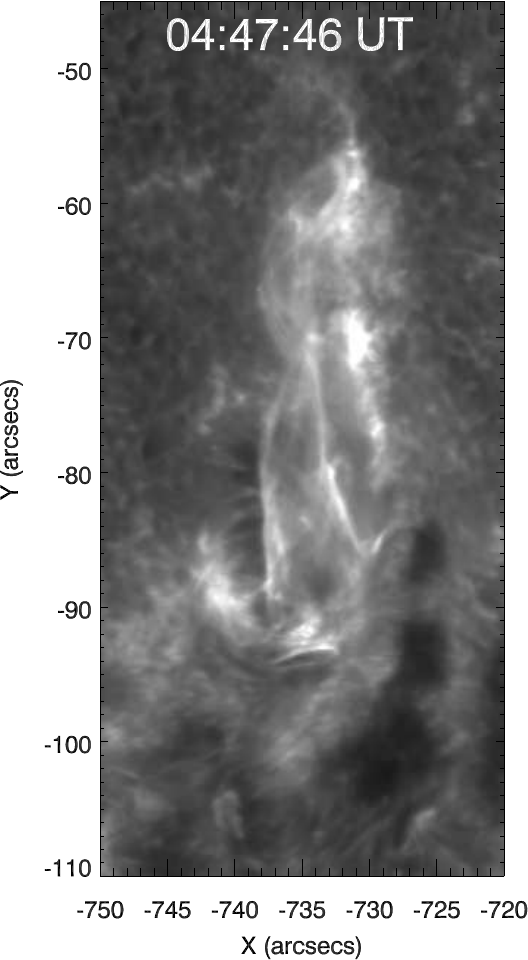}
\includegraphics[width=2.75cm]{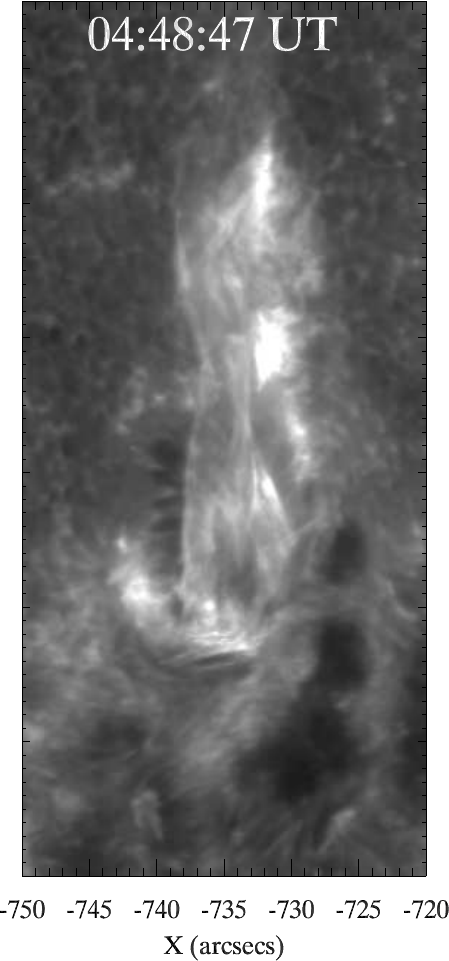}
\includegraphics[width=2.75cm]{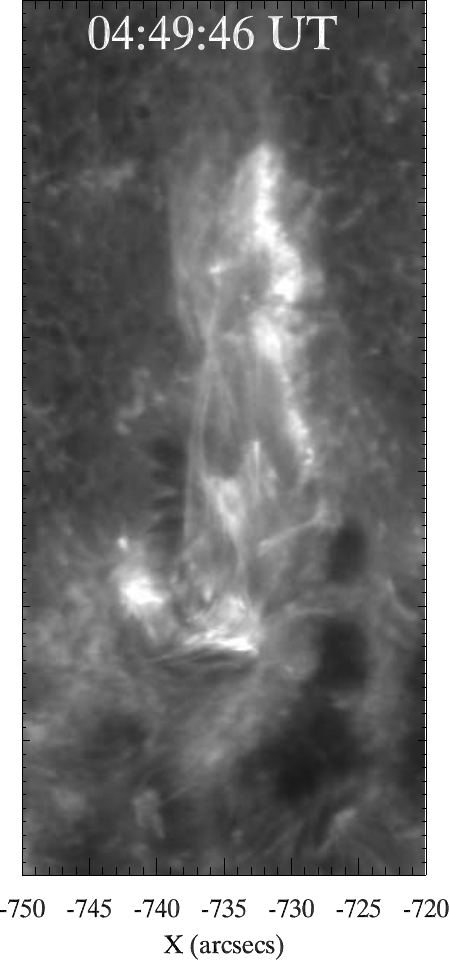}
\includegraphics[width=2.75cm]{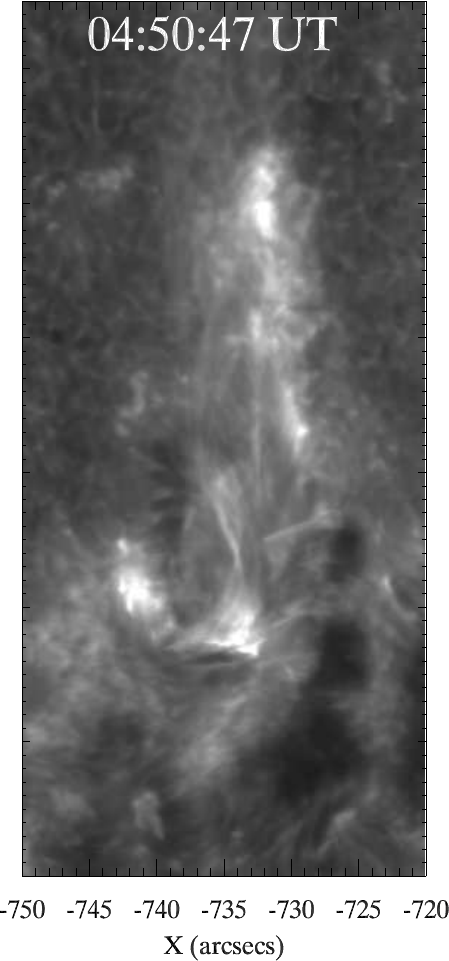}
\includegraphics[width=2.75cm]{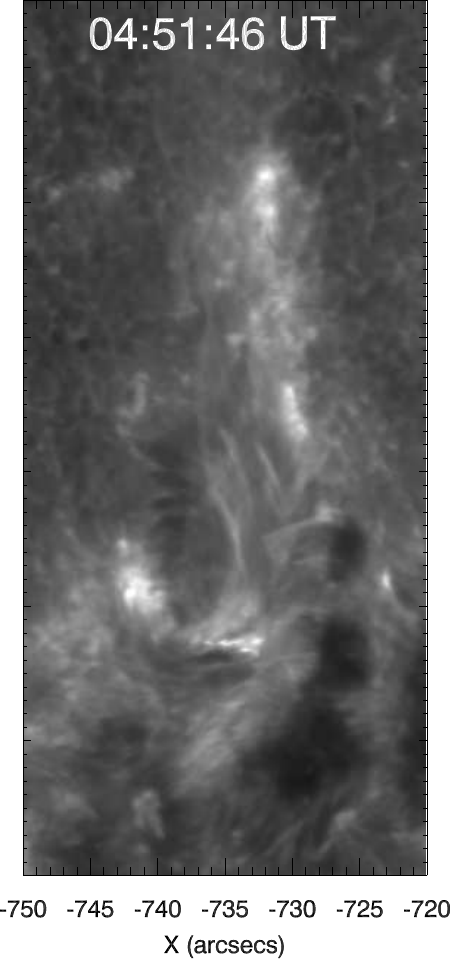}
}

\caption{Time sequence of SOT/Ca II H (3968 \AA ) images during
04:43 UT and 04:52 UT at 04 June, 2007. The images show
high-resolution chromospheric morphology of flaring loop. Activation
of helical twist and simultaneous brightening in the chromosphere
are clearly seen during B5.0 class flare (04:40 UT--04:51 UT).}
\label{fig1}
\end{figure*}
\clearpage
\begin{figure*}
{
\hspace{0.1cm}
\includegraphics[width=3.35cm]{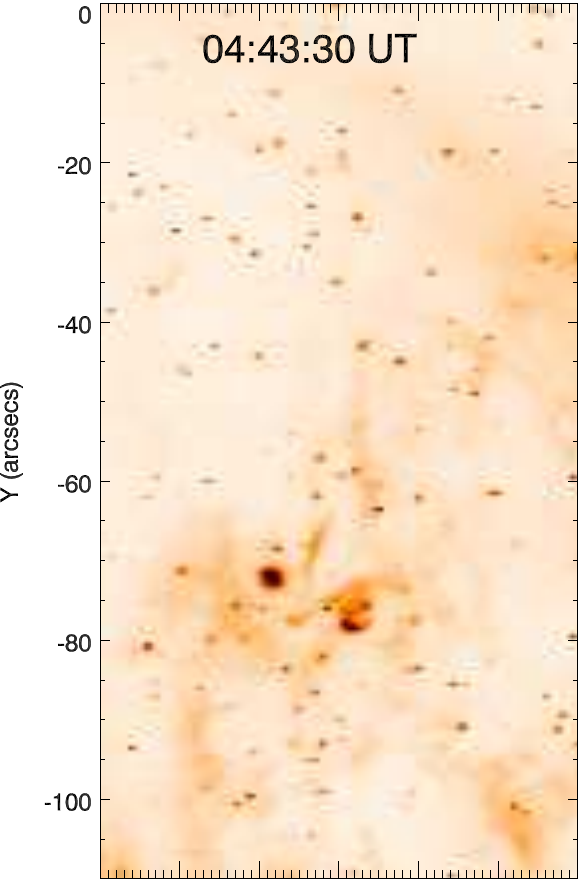}
\hspace{0.12cm}
\includegraphics[width=2.8cm]{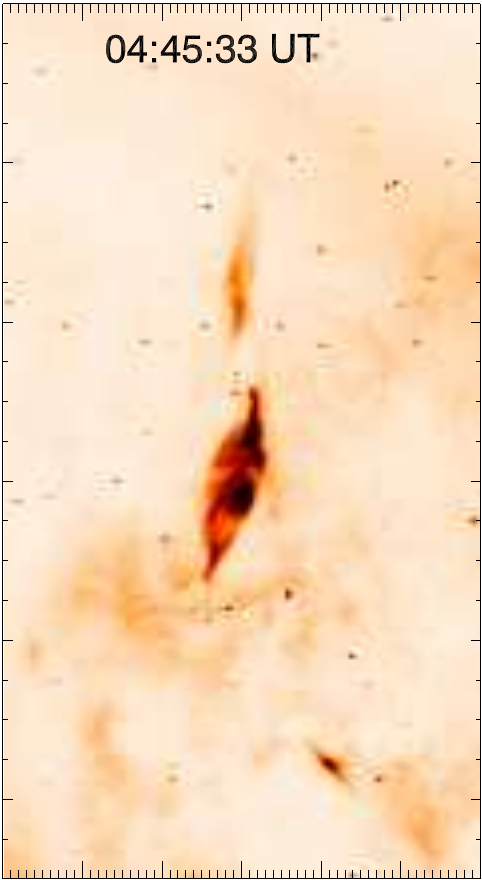}
\hspace{0.09cm}
\includegraphics[width=2.8cm]{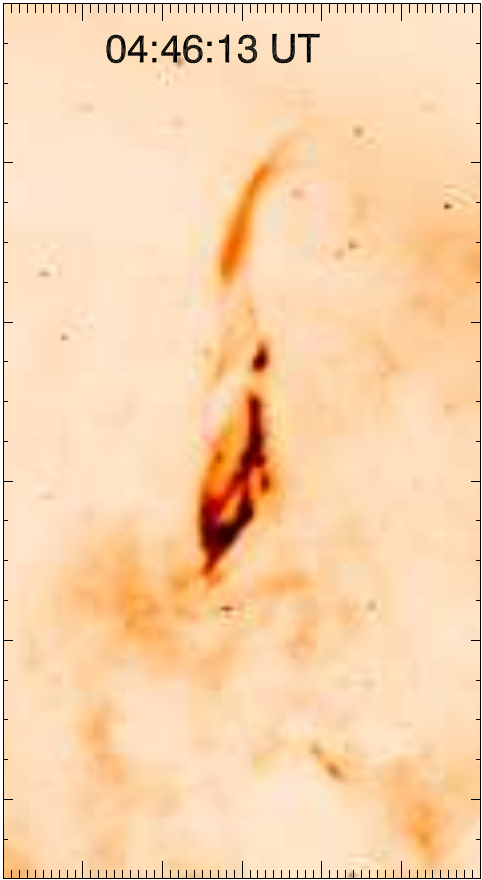}
\hspace{0.09cm}
\includegraphics[width=2.8cm]{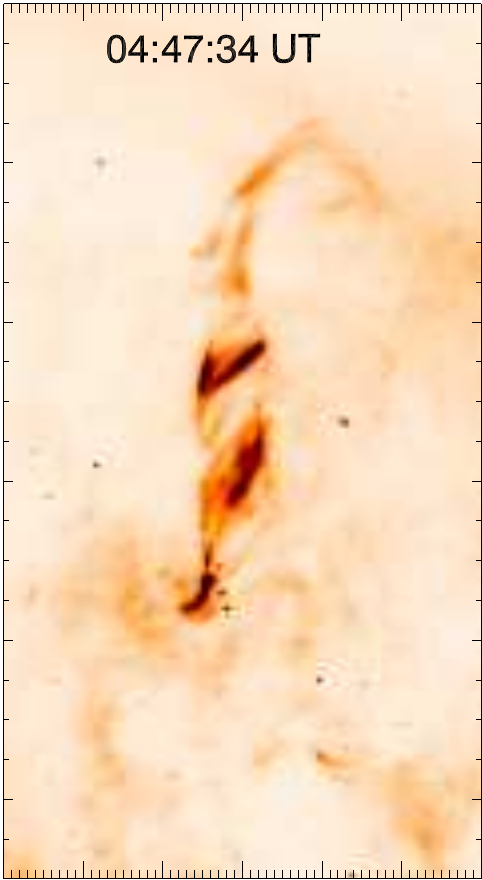}
\hspace{0.09cm}
\includegraphics[width=2.8cm]{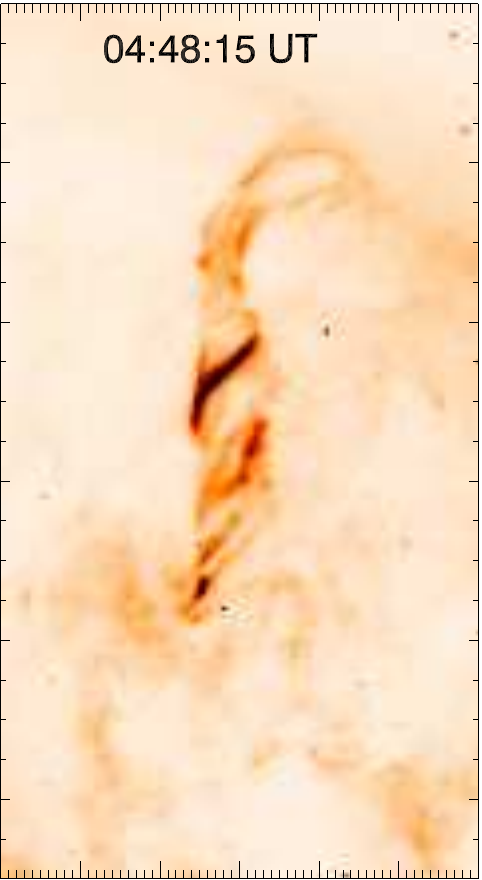}

\hspace{0.1cm}
\includegraphics[width=3.5cm]{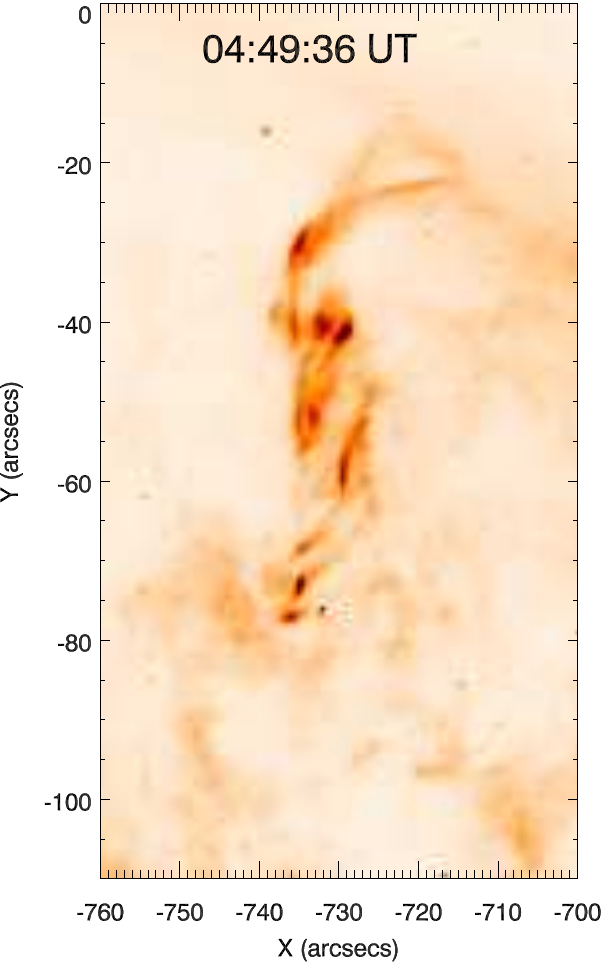}
\includegraphics[width=3.05cm]{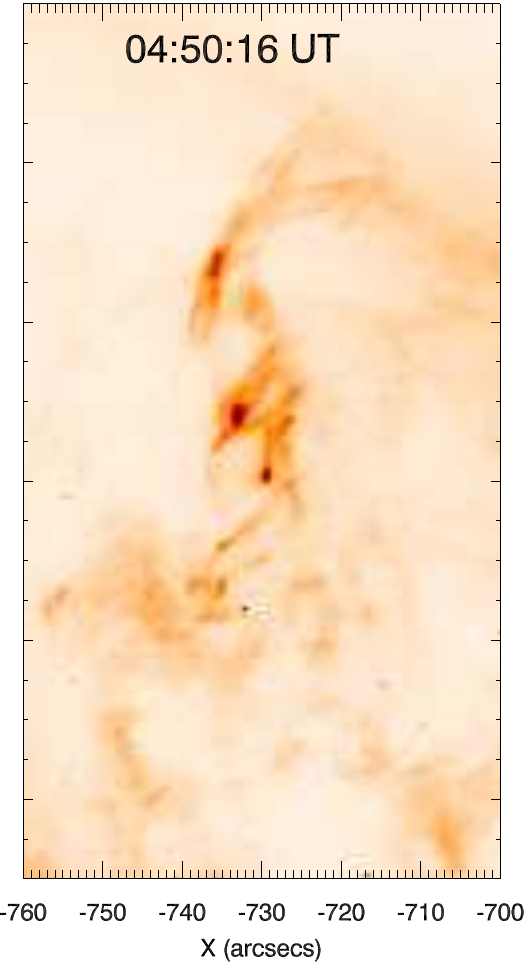}
\includegraphics[width=3.05cm]{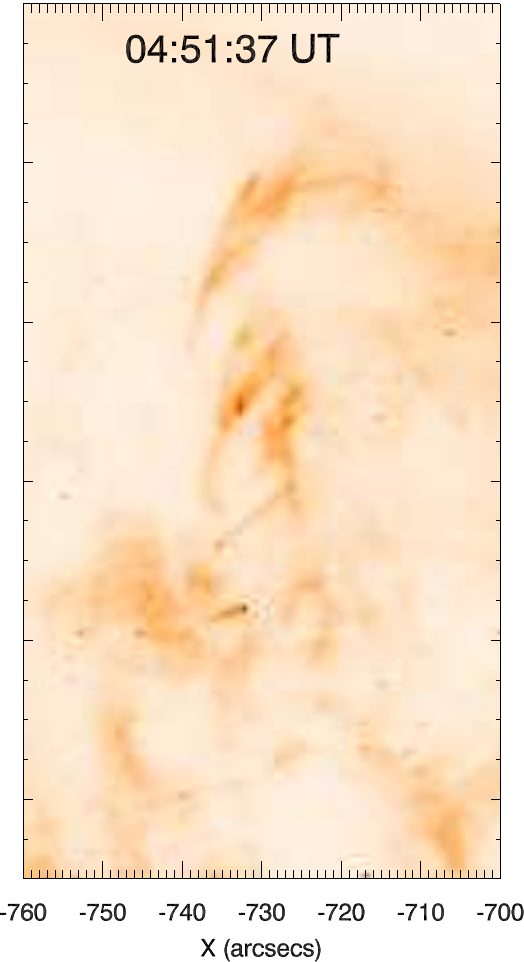}
\includegraphics[width=3.05cm]{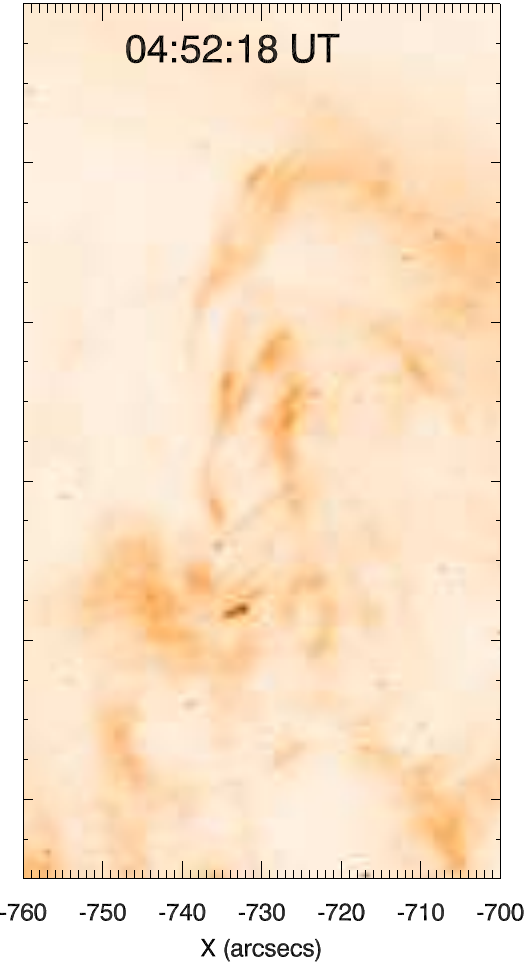}
\hspace{0.01cm}
\includegraphics[width=3.05cm]{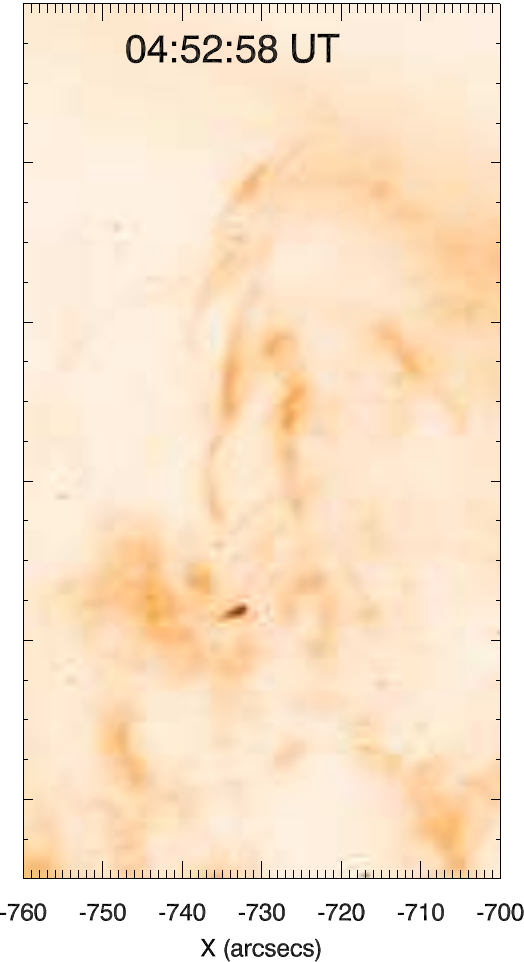}
} \caption{Time sequence of TRACE 171 \AA \ Fe IX images of flaring
loop in the AR 10960 during 04:43 UT and 04:52 UT on 04 June, 2007.
The images are in reverse color and show clear helical twist of the
loop during the B5.0 flare. Note the double structure of coronal loop
top between 04:47-04:51 UT near (X,Y)=(-720,-20).} \label{fig1}
\end{figure*}

\end{document}